\documentclass{elsart}
\usepackage{graphicx}

\begin{document}

\begin{frontmatter}

\title{\bf Anisotropic fluid dynamics in the early stage
of relativistic heavy-ion collisions \thanksref{grant}} 
\thanks[grant]{e-mail: Wojciech.Florkowski@ifj.edu.pl \\
Supported in part by the Polish Ministry of Science and Higher Education,
grant No. N202 034 32/0918.}

\author[ifj,ujk]{Wojciech Florkowski} 

\address[ifj]{The H. Niewodnicza\'nski Institute of Nuclear Physics, Polish Academy of Sciences, ul. Radzikowskiego 152, 
31-342 Krak\'ow, Poland}

\address[ujk]{Institute of Physics, Jan Kochanowski University, 
ul.~\'Swi\c{e}tokrzyska 15, 25-406~Kielce, Poland} 

\begin{abstract} 
A formalism for anisotropic fluid dynamics is proposed. It is designed to describe boost-invariant systems with anisotropic pressure. Such systems are expected to be produced at the early stages of relativistic heavy-ion collisions, when the timescales are too short to achieve equal thermalization of transverse and longitudinal degrees of freedom. The approach is based on the energy-momentum and entropy conservation laws, and may be regarded as a minimal extension of the boost-invariant standard relativistic hydrodynamics of the perfect fluid. We show how the formalism may be used to describe the isotropization of the system (the transition from the initial state with no longitudinal pressure to the final state with equal
longitudinal and transverse pressure).
\end{abstract}
\end{frontmatter}
\vspace{-7mm} PACS: 25.75.Dw, 25.75.-q, 21.65.+f, 14.40.-n

Keywords: hydrodynamic evolution, transversal equilibrium, heavy-ion collisions 

{\bf 1.}  At present, the most successful description of early parton dynamics is achieved with the help of the relativistic hydrodynamics \cite{Kolb:2003dz,Huovinen:2003fa,Shuryak:2004cy}. With the equation of state incorporating the phase transition and with the appropriate modeling of the freeze-out process, the hydrodynamic approach leads to very successful description of the hadron transverse-momentum spectra and the elliptic flow coefficient $v_2$ \cite{teaney,hiranoetal,nonaka}. We also note that with a suitable modification of the initial conditions, the hydrodynamic approach describes consistently the HBT radii \cite{Broniowski:2008vp}. 

In spite of those clear successes, the use of the hydrodynamics is faced with the problem of so-called early thermalization –-- in order to have a successful description of the data, the hydrodynamic evolution (implicitly assuming the three-dimensional local equilibrium) should start at a very early time, well below 1 fm/c. Such short values can be hardly explained within the microscopic calculation. 

Recently, a possible solution to the problem of early thermalization has been proposed \cite{Bialas:2007gn}. With the assumptions that only transverse degrees of freedom are thermalized and the longitudinal dynamics is essentially the free-streaming, one can obtain the parton transverse-momentum spectra and $v_2$ which agree well with the data \cite{Bialas:2007gn,Chojnacki:2007fi}. In this approach, called below the {\it transverse hydrodynamics} (for general formulation see \cite{Ryblewski:2008fx}), the longitudinal pressure vanishes while the transverse pressure is large and leads to the formation of a substantial transverse flow, which is the main effect responsible for the good agreement with the data. 

One naturally expects that after some time the purely transverse hydrodynamic evolution is transformed into the standard hydrodynamic evolution with isotropic pressure. The typical way to describe such transformation would be to use the kinetic theory \cite{Zhang:2008kj} or dissipative hydrodynamics \cite{Kovchegov:2005az,Bozek:2007di}. Such approaches lead, however, to the entropy and particle production which seems to contradict the observed scalings of the hadron production with the number of the initial constituents (unless the entropy production is a small effect). Another mechanism to describe a similar transition, from the initial quasithermal two-dimensional parton distribution to the final three-dimensionally isotropic parton distribution, was studied in Ref. \cite{Dumitru:2005gp}. This approach conserves the entropy and is based on the coupled Vlasov and Yang-Mills equations for the quark-gluon plasma. The full isotropization of the system is in this case an effect of bending of parton trajectories in strong color fields.

The aim of this paper is to propose the extension of the standard boost-invariant hydrodynamics, which would be suitable for the description of systems with anisotropic pressure. In the special cases, our approach is reduced to the transverse hydrodynamics or standard hydrodynamics. It may be also used to describe {\it effectively} the process of full isotropization of the pressure. 

The proposed formalism is based on the energy, momentum, and entropy conservation laws. As we have mentioned above, the use of the entropy conservation is suggested by various modeling of the RHIC data. For example, the PHOBOS data \cite{Back:2004mr} shows that the numbers of produced hadrons per wounded nucleon in central d+Au and Au+Au collisions are very much similar (differences smaller than 30\%). This indicates that  equilibration/isotropization effects do not produce a large amount of the entropy. In addition, the recently observed scalings of the hadron production with the number of so-called wounded constituents \cite{Bialas:2007eg} leave little room for extra particle production during the evolution of the system.

{\bf 2. } Our starting point is the following form of the energy-momentum tensor~\footnote{Throughout the paper we use the natural units with $c = \hbar = 1$. The metric tensor $g^{\mu \nu} = \hbox{diag}(1,-1,-1,-1)$.}
\begin{equation}
T^{\mu \nu} = \left(\varepsilon  + P_T \right) U^{\mu}U^{\nu} 
- P_T g^{\mu\nu} - (P_T-P_L) V^{\mu}V^{\nu},
\label{tensorT1}
\end{equation}
where $\varepsilon$ is the energy density, $P_T$ and $P_L$ are the transverse and longitudinal pressure, and $U^\mu$ is the four-velocity of the fluid satisfying the normalization condition $U^\mu U_\mu = 1$. In the isotropic case, the pressures $P_T$ and $P_L$ are equal, $P_T = P_L = P$, and the energy-momentum tensor takes the standard form. For the anisotropic fluid, the last term in (\ref{tensorT1}) is different from zero. The fourvector $V^\mu$ defines the direction of the longitudinal pressure. It is spacelike, orthogonal to $U^\mu$, $U^\mu V_\mu = 0$, and normalized by the condition $V^\mu V_\mu = -1$. In the local rest frame of the fluid element we have $U^\mu = (1, 0, 0, 0)$ and $V^\mu = (0, 0, 0, 1)$, hence the energy-momentum tensor (\ref{tensorT1}) takes the expected form
\begin{equation}
T^{\mu \nu} =  \left(
\begin{array}{cccc}
\varepsilon  & 0 & 0 & 0 \\
0 & P_T & 0 & 0 \\
0 & 0 & P_T & 0 \\
0 & 0 & 0 & P_L
\end{array} \right).
\end{equation}
For the boost-invariant systems the fluid four-velocity $U^\mu$ has the structure
\begin{equation}
U^\mu = \left(u^0 \cosh\eta,u_x,u_y, u^0 \sinh\eta \right), 
\label{bigu}
\end{equation}
where $u^\mu = (u^0, u_x, u_y, 0)$ is the fluid four-velocity at $z = 0$ (for the vanishing longitudinal coordinate), and $\eta = 1/2 \ln(t + z)/(t - z)$ is the spacetime rapidity. With the normalization condition $u_0^2 - u_x^2 - u_y^2 = 1$ we automatically have $U^{\,2} = 1$. The four-vector $V^\mu$ satisfying the appropriate normalization and orthogonality conditions has the form
\begin{equation}
V^\mu = \left(\sinh\eta, 0, 0,  \cosh\eta \right). 
\label{bigV}
\end{equation}
In this paper we restrict our considerations to the case of massless particles,
where $T^\mu_{\,\,\,\mu} = 0$ and $\varepsilon = 2 P_T + P_L$. As in the standard hydrodynamics, the evolution equations are obtained from the energy-momentum conservation
law
\begin{equation}
\partial_\mu T^{\mu \nu}=0.
\label{emcl}
\end{equation}

The projection of Eq. (\ref{emcl}) on the four-velocity $U_\nu$ yields
\begin{equation}
D \varepsilon + (\varepsilon + P_T) \partial_\mu U^\mu 
- \Delta U_\nu V^\mu \partial_\mu V^\nu = 0.
\label{Deps1}
\end{equation}
where we have introduced the short-hand notation for the total time derivative, \mbox{$D \equiv U^\mu \partial_\mu $} and the difference of the pressures, $\Delta = P_T - P_L$. In addition to the energy-momentum conservation law (\ref{emcl}) we demand that there is a conserved entropy current characterizing the system. We write it in the form $\partial_\mu (\sigma U^\mu) = 0$ or equivalently as
\begin{equation}
D \sigma + \sigma \partial_\mu U^\mu  = 0.
\label{entr-cons2}
\end{equation}
where $\sigma$ is the entropy density. Moreover, with the definition (\ref{bigV}) one finds 
\begin{equation}
V^\mu \partial_\mu V^\nu = \partial^\nu \ln \tau,
\label{funtau}
\end{equation}
where $\tau = \sqrt{t^2 - z^2}$ is the longitudinal proper time. Eqs. (\ref{entr-cons2}) and (\ref{funtau}) allow us to rewrite Eq. (\ref{Deps1}) in the form
\begin{equation}
D \varepsilon = \frac{(\varepsilon + P_T)}{\sigma} D \sigma
+ \frac{\Delta}{\tau} D \tau.
\label{Deps2}
\end{equation}
Eq. (\ref{Deps2}) indicates that in the general case the energy density may be considered as a function of the two variables,     $\varepsilon=\varepsilon(\sigma,\tau)$, hence $\varepsilon(\tau,x,y) = \varepsilon\left[\sigma(\tau,x,y), \tau \right]$. We emphasize that this is a novel feature of our approach, which distinguishes it from the standard hydrodynamics where the energy density depends only on the entropy density, $\varepsilon=\varepsilon(\sigma)$ and $\varepsilon(\tau,x,y) = \varepsilon\left[\sigma(\tau,x,y)\right]$.

{\bf 3.} The functional dependence $\varepsilon = \varepsilon(\sigma,\tau)$ plays a role of the {\it generalized equation of state} for our system. Eq. (\ref{Deps2}) is satisfied if the two conditions hold,
\begin{equation}
\left(\frac{\partial \varepsilon}{\partial \sigma} \right)_\tau= \frac{\varepsilon+P_T}{\sigma}, \quad
\left( \frac{\partial \varepsilon}{\partial \tau} \right)_\sigma = \frac{\Delta}{\tau}.
\label{thermid1}
\end{equation} 
Multiplying Eqs. (\ref{thermid1}) by $\sigma$ and $\tau$, respectively, and dividing both of them by $\varepsilon$ we obtain a simple set of equations
\begin{equation}
\left(\frac{\partial \varepsilon^\prime}{\partial \sigma^\prime} \right)_{\tau^\prime}= 
\frac{4}{3}+\frac{\Delta^\prime}{3}, \quad
\left( \frac{\partial \varepsilon^\prime}{\partial \tau^\prime} \right)_{\sigma^\prime} = 
\Delta^\prime,
\label{thermid2}
\end{equation}
where $\varepsilon^\prime = \ln(\varepsilon/\varepsilon_0)$, 
$\Delta^\prime = \Delta/\varepsilon$, $\sigma^\prime = \ln(\sigma/\sigma_0)$, and $\tau^\prime = \ln(\tau/\tau_0)$ with $\varepsilon_0,\sigma_0, \tau_0$ being arbitrary constants \footnote{Their values may be arranged to impose the appropriate initial conditions.}. The thermodynamic consistency requires that $d\varepsilon^\prime$ is the total derivative, hence the mixed second derivatives of the function $\varepsilon^\prime(\sigma^\prime,\tau^\prime)$ are equal. This condition implies that $\Delta^\prime$ must be a function of the single variable, namely $\Delta^\prime = \Delta^\prime(\sigma^\prime + 3\tau^\prime)$. By the direct integration of Eq. (\ref{thermid2}) we find that the energy density function must be of the form 
\begin{equation}
\varepsilon=\varepsilon_0 \left( \frac{\sigma}{\sigma_0} \right)^{4/3}
R\left(x\right), \quad x = \frac{\sigma}{\sigma_0} \left( \frac{\tau}{\tau_0} \right)^3.
\label{generaleps}
\end{equation}
The function $R(x)$ is related to the function $\Delta^\prime(x)$ by the equation
\begin{equation}
R(x) = \exp \left[ \frac{1}{3} \int\limits_0^{\ln x}
\Delta^\prime(y) dy \right].
\label{RDelta}
\end{equation}
In the similar way we may express the transverse and longitudinal pressure, namely
\begin{equation}
P_T = \varepsilon_0 \left( \frac{\sigma}{\sigma_0} \right)^{4/3}
\left[ \frac{R}{3} + x R^\prime \right], \quad 
P_L = \varepsilon_0 \left( \frac{\sigma}{\sigma_0} \right)^{4/3}
\left[ \frac{R}{3} - 2 x R^\prime \right]
\label{PTandPL}
\end{equation}
where $R^\prime = dR/dx$. We note that Eqs. (\ref{generaleps}) and (\ref{PTandPL}) satisfy explicitly the condition $\varepsilon = 2 P_T + P_L$.

Equation (\ref{RDelta}) indicates that $R$ depends on the space-time evolution of the pressure anisotropy. As a consequence, its particular form may be obtained only from the microscopic calculation. At the hydrodynamic level of description, which is the subject of our analysis, the form of the function $R$ should be assumed. Nevertheless, from the constraints $P_T \geq P_L$ and $P_L \geq 0$, we obtain the two conditions: $R^\prime(x) \geq 0$ and $R(x) \geq 6 x R^\prime(x)$, hence $R(x)$ cannot grow faster than $x^{1/6}$. Moreover, we expect that $R$ saturates for large values of time, when the transition to the standard hydrodynamics takes place. Below we shall refer to $R$ as to the {\it pressure relaxation function}. With a given form of $R$, the generalized equation of state (\ref{generaleps}) and also the dynamics of the system are well defined.

We define the temperature in the standard way as the derivative of the energy density with respect to the entropy density,
\begin{equation}
T = \left(\frac{\partial \varepsilon}{\partial \sigma}\right)_\tau,
\label{temp}
\end{equation}
which together with Eq. (\ref{Deps2}) leads to the following {\it generalized thermodynamic identities}
\begin{equation}
T \sigma = \varepsilon + P_T, \quad 
D \varepsilon = T D\sigma + \Delta D \ln \tau, \quad
D P_T = \sigma DT - \Delta D \ln \tau.
\label{genthermid}
\end{equation}
One can notice that for the isotropic system, where $\Delta = 0$ and $P_T = P_L = P$, the above equations are reduced to the standard thermodynamic identities valid for the systems with zero baryon density.

{\bf 4.} Starting from the energy-momentum conservation laws (\ref{emcl}) and using Eqs. (\ref{genthermid}) we obtain the final form of the hydrodynamic equations for the anisotropic perfect fluid:
\begin{eqnarray}
U^\mu \partial_\mu \left(T U^\nu \right) &=& \partial^\nu T, \nonumber \\
\partial_\mu \left( \sigma U^\mu \right) &=& 0. 
\label{hydro-eq}
\end{eqnarray}
In the derivation of Eqs. (\ref{hydro-eq}), which are the main formal result of our paper,  we have used the following properties: $V^\mu \partial_\mu \tau = 0$, $\partial_\mu V^\mu = 0$, and \mbox{$V^\mu \partial_\mu \Delta = 0$}. They follow directly from the definition (\ref{bigV}) and the assumption of  boost-invariance. Since the projection of the first equation in (\ref{hydro-eq}) on $U_\nu$ yields identically zero and, in addition, the system is boost-invariant, Eqs. (\ref{hydro-eq}) consist of 3 equations for 4 unknown functions: $u_x$, $u_y$, $T$, and $\sigma$. Similarly as in the standard hydrodynamics, the "dynamical" Eqs. (\ref{hydro-eq}) must be supplied with the "material" equation --- the equation of state in the standard hydrodynamics or the generalized equation of state of the form (\ref{generaleps}) in our approach. It is very interesting to observe that Eqs. (\ref{hydro-eq}) have exactly the same form as the standard hydrodynamic equations for boost-invariant baryon-free matter. The only difference remains in the use of the generalized equation of state including the pressure relaxation function $R$. The latter defines the temperature $T$ as a function of $\sigma$ and $\tau$. Equivalently, it may be used to define $\sigma$ as a function of $T$ and $\tau$.

{\bf 5.} Let us now consider a few special cases. To illustrate the main characteristic features of our approach with simple and analytic examples, we neglect the transverse expansion of the fluid. In this case, the entropy equation has always the solution of the form
\begin{equation}
\sigma\left[ T(\tau),\tau \right] = \sigma_0 \frac{\tau_0}{\tau}.
\end{equation}
{\bf i)} The isotropic case with $P_T = P_L$ corresponds to the trivial choice $\Delta^\prime = 0$ and gives $R(x) = 1$. The generalized equation of state in this case is reduced to the well known black-body equation $\varepsilon = \varepsilon_0 (\sigma/\sigma_0)^{4/3}$, which yields $\sigma \sim T^3$. In this way we recover the famous result of the Bjorken model where $T \sim \tau^{-1/3}$.

{\bf ii)} Another interesting situation corresponds to the case with vanishing longitudinal pressure. For $P_L = 0$ one finds $\Delta^\prime = 1/2$, which in turn leads to the relation $R(x) = x^{1/6}$. In this case the generalized equation of state gives $\sigma \sim T^{\,2}/\tau$ \cite{Bialas:2007gn,Ryblewski:2008fx}. Comparing this result to Eq. (18) we conclude that the temperature becomes independent of time. This is an expected result of transverse hydrodynamics –-- without transverse expansion the transverse clusters cannot cool down and their temperature remains constant. 

{\bf iii)} As the third case w consider the pressure relaxation function of the form $R(x) = [x_I - (x_I - 1) \exp[-(x - 1)/(x_I - 1)]]^{1/6}$. Having in mind that the variable $x$ grows with time, $x = (\sigma/\sigma_0) (\tau/\tau_0)^3 = (\tau/\tau_0)^2$, this form of the function $R$ may be useful to describe a {\it smooth isotropization} transition from two-dimensional purely transverse hydrodynamic expansion to the isotropic expansion governed by the standard hydrodynamics. The parameter $x_I$ determines here the timescale $\tau_I$ for the isotropization transition, $\tau_I^2 = x_I \tau_0^2$. For $x -1 \ll  x_I - 1$ we have $R \approx x^{1/6}$, hence we recover the conditions of transverse hydrodynamics. On the other hand, for $x -1 \gg x_I - 1$ the function $R$ becomes a constant, $R = x_I^{1/6}$, and we recover the conditions of standard hydrodynamics, however, with a rescaled initial energy density \footnote{If the system approaches the limit of standard hydrodynamics with local thermal equilibrium, the combination of the constants $x_I^{1/6} \varepsilon_0 \sigma_0^{-4/3}$ must be expressed by the number of the internal degrees of freedom. This condition leads to the constraint for $\varepsilon_0$ and $\sigma_0$, which are the initial parameters for the transverse hydrodynamics. The constraint may be easily fulfilled since for the transverse hydrodynamics the combination  $\varepsilon \sigma^{-4/3}$ is not a constant \cite{Ryblewski:2008fx}.}.

{\bf iv)} Finally, we consider the case $R(x) = x^{1/6} \theta(x_I - x) + x_I^{1/6} \theta(x - x_I)$, where $\theta(x)$ is the step function. Similarly to the previous case, the parameter $x_I$ determines the timescale for the isotropization transition. However, in this case we deal with a {\it sudden isotropization} transition. Certainly, this kind of the transition may be treated only as an approximation to real processes. 

\begin{figure}[t]
\begin{center}
\includegraphics[angle=0,width=0.75\textwidth]{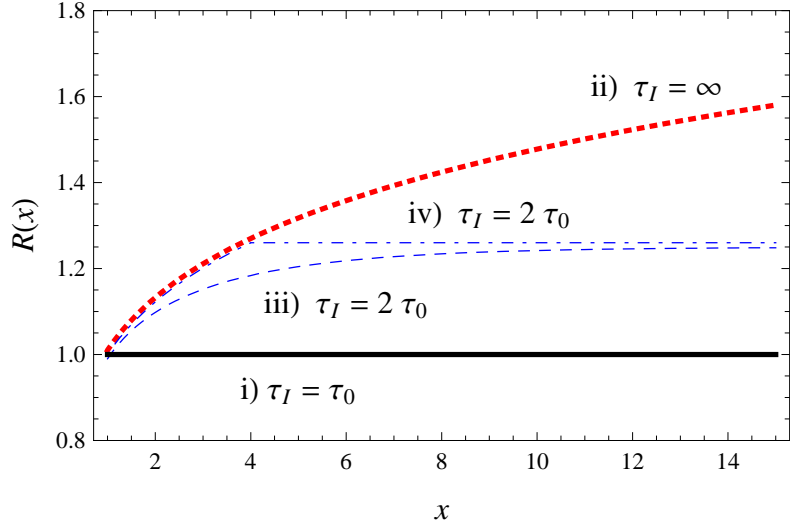}
\end{center}
\caption{\small {Four examples of the pressure relaxation function $R(x)$ discussed in the text in the cases i) - iv).} }
\label{fig1}
\end{figure}

In Fig. 1 we show four examples of the function $R$, introduced above in the cases i) - iv). The solid line describes the standard hydrodynamics (this type of the evolution may be considered as the case with $\tau_I = \tau_0$, i.e., the isotropization takes place right at the beginning of the evolution), the dotted line corresponds to the purely transverse hydrodynamics ($\tau_I = \infty$, the isotropization never happens), while the dashed and dashed-dotted lines correspond to the forms discussed in the cases iii) and iv) --- the smooth and sudden isotropization with $\tau_I = 2 \tau_0$. 

In Fig. 2 we show the time evolution of the energy density determined by the relaxation functions shown in Fig. 1 (with the same notation). The solid line describes the case of the standard 1+1 boost-invariant hydrodynamics where $\varepsilon \sim \tau^{-4/3}$, whereas the dotted line describes the case of the transverse hydrodynamics where $\varepsilon \sim \tau^{-1}$. The dashed and the dashed-dotted lines describe the situations with a transition from transverse to standard hydrodynamics.  We observe that
the more the evolution resembles that of transverse hydrodynamics, the slower is 
the decrease of the energy density. This is clearly caused by the fact that no longitudinal work is done in the case of the transverse hydrodynamics, hence less energy is "lost" in this case. 

\begin{figure}[t]
\begin{center}
\includegraphics[angle=0,width=0.75\textwidth]{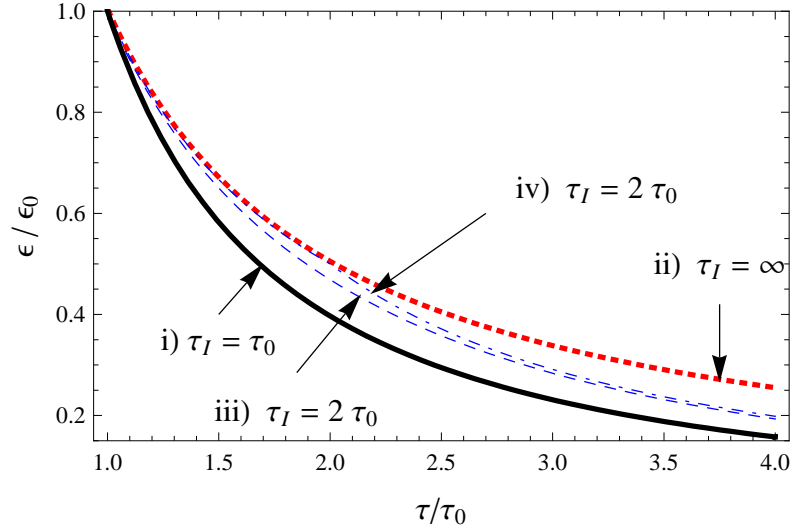}
\end{center}
\caption{\small { Time evolution of the energy-density in the cases i) - iv).} }
\label{fig2}
\end{figure}

\begin{figure}[t]
\begin{center}
\includegraphics[angle=0,width=0.75\textwidth]{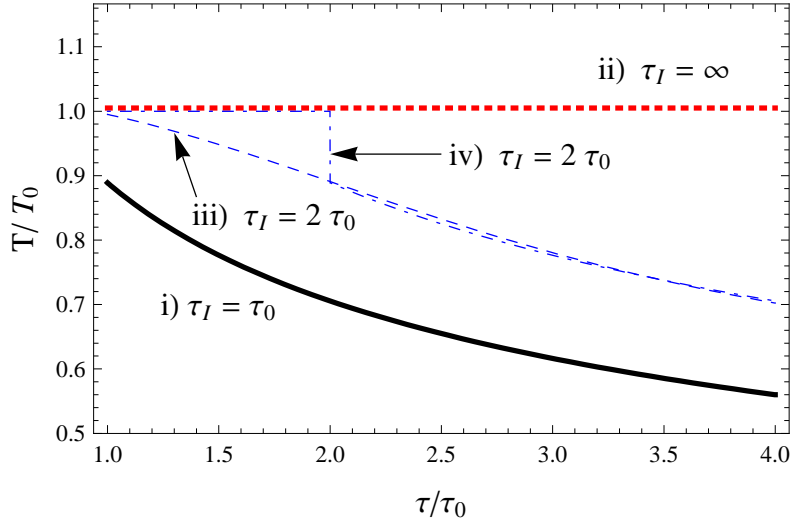}
\end{center}
\caption{\small {Time evolution of the temperature for the cases i) --- iv).  } }
\label{fig3}
\end{figure}

In Fig. 3 we describe the temperature evolution. Again the solid line describes the standard 1+1 boost-invariant hydrodynamics where $T \sim \tau^{-1/3}$, while the dotted curve represents the case of transverse hydrodynamics where \mbox{$T$ = const}~\footnote{The difference at the beginning of the two curves is caused by our normalization which assumes the same energy density at $\tau=\tau_0$}. Interestingly, in the case with the sudden isotropization the temperature exhibits a discontinuity. This behaviour is caused by the fact that the longitudinal degrees of freedom are suddenly "switched on" .  Consequently, the energy and entropy must be shared among larger number of degrees of freedom and the temperature decreases (note that this behavior is a consequence of the assumptions on the energy and entropy conservations).

{\bf 6.}  We close the paper with the following conclusions:

i) Our main formal result is the set of the hydrodynamic equations (\ref{hydro-eq}) supplied with the generalized equation of state (\ref{generaleps}). These equations determine the time evolution of boost-invariant perfect fluids with anisotropic pressure. The generalized equation of state includes the pressure relaxation function $R$, which may be used to define the way in which the system reaches complete isotropization. 

We note that although the examples given in this paper illustrated the 1+1 boost-invariant expansion, Eqs. (\ref{hydro-eq}) are suitable for the numerical analysis of the 3+1 boost-invariant systems.  

ii) The use of the proposed formalism in the phenomenological approaches to relativistic heavy-ion collisions is straightforward and may bring interesting results. For example, the presence of the initial purely transverse hydrodynamic stage leads to the formation of a very strong flow \cite{Bialas:2007gn}. As it has been shown recently in Ref. [4], the presence of the strong flow is one of the necessary conditions required to describe uniformly the RHIC soft hadronic data.

iii) The internal consistency of the approach requires that the pressure relaxation function $R(x)$ does not grow faster than $x^{1/6}$, and saturates for large arguments (in the cases when the evolution is sufficiently long and the system becomes isotropic). The fitting of the heavy-ion data has the potential of determining the relaxation function $R$, whose behavior is determined essentially by one parameter –-- the timescale for the isotropization transition. 

iv) The entropy conservation assumed in our approach suggests that our description of the full isotropization may be an effective way of dealing with the effects caused by the mean-field quark-gluon plasma dynamics, based on the entropy-conserving collisionless kinetic equations \cite{Arnold:2004ti,Dumitru:2005gp,Mrowczynski:2005gw}. This issue deserves further studies. Similarly, the generalizations of our approach to non boost-invariant cases represents an interesting and open problem.

{\it Acknowledgements:} I thank Professor Kacper Zalewski for the critical reading of the manuscript and interesting comments. I also thank Piotr Bozek, Wojtek Broniowski and Giorgio Torrieri for discussions and critical remarks. 

%\bibliography{ref-rr}

\end{document}